\documentclass[
    10pt,
    superscriptaddress,
    nofootinbib,
    amsfonts,
    amssymb,
    twocolumn,
    aps,
    pra,
    longbibliography
]{revtex4-2}

\usepackage[utf8]{inputenc} % set input encoding UTF 8 bit
\usepackage[T1]{fontenc} % use T1 for most European languages
\usepackage[english]{babel} % translate document elements
\usepackage{lipsum} % generate random text

\usepackage{geometry} % define the page structure
\usepackage{enumitem} % manage enumerating and itemize
\usepackage{comment} % selectively include document pieces
\usepackage[normalem]{ulem} % provide \ul (underline) command
\usepackage{soul} % provide \st for striking out, \ul for underlining, etc.
\setstcolor{red}

\usepackage{newfloat} % manage interface for floating objects

\usepackage{graphicx,xcolor}
\usepackage{tikz} % creates graphic elements in LaTeX
\usetikzlibrary{arrows.meta}
\usetikzlibrary{positioning}

\usepackage[edges]{forest} % drawing blocks trees
\usepackage[most]{tcolorbox} % create color boxes

\usepackage{amsmath} % provide math expressions
\usepackage{amsthm} % define definitions and theorems

\usepackage{eucal} % provide \mathcal letters
\usepackage{dsfont} % provide \mathds letters

\usepackage{mathtools} % define equal for eq. package
\usepackage{physics} % provide Dirac's brackets

\definecolor{colPaW}{cmyk}{0.80, 0.13, 0.14, 0.04, 1.00}
\definecolor{colGLM}{rgb}{0.86, 0.82, 1.0}
\definecolor{colQub}{rgb}{1.0, 0.65, 0.0}
\definecolor{colSem}{rgb}{0.82, 0.41, 0.12}
\definecolor{colPos}{cmyk}{0, 0.76, 0.33, 0.02}

\makeatletter
\def\blfootnote{\xdef\@thefnmark{}\@footnotetext}
\makeatother

\makeatletter
\def\hamilt{\CMcal{\hat{H}}}
\def\iden{\mathds{1}}
\def\hilbert{\mathcal{H}}
\def\observable{\hat{\mathcal{O}}}
\makeatother

\begin{document}

\title{Quantum Time and the Time-Dilation induced\\ Interaction Transfer mechanism}

\author{Dario Cafasso}
\email{email: cafasso.dario@gmail.com}
\affiliation{Department of Physics, University of Pisa, and INFN - Sezione di Pisa, Polo Fibonacci, Largo B. Pontecorvo 3, IT-56127 Pisa, Italy}

\author{Nicola Pranzini}
\affiliation{QTF Centre of Excellence and Department of Physics, University of Helsinki, FIN-00014 Helsinki, Finland}
\affiliation{InstituteQ - the Finnish Quantum Institute, Finland}

\author{Jorge Yago Malo}
\affiliation{Department of Physics, University of Pisa, and INFN - Sezione di Pisa, Polo Fibonacci, Largo B. Pontecorvo 3, IT-56127 Pisa, Italy}

\author{Vittorio Giovannetti}
\affiliation{NEST, Scuola Normale Superiore and Istituto di Nanoscienze,
Consiglio Nazionale delle Ricerche, Piazza dei Cavalieri 7, IT-56126 Pisa, Italy}

\author{Marilù Chiofalo}
\affiliation{Department of Physics, University of Pisa, and INFN - Sezione di Pisa, Polo Fibonacci, Largo B. Pontecorvo 3, IT-56127 Pisa, Italy}

\begin{abstract}
Given a bipartite quantum system in an energy eigenstate, the dynamical description for one component can be derived via entanglement using the other component as a clock. This is the essence of the Page and Wootters mechanism. 
Moreover, if the clock is subject to a gravitational-like interaction,
relative time evolution is then described by a \textit{Time-Dilated Schrödinger equation}, in which the so-called \textit{redshift operator} describes a purely quantum effect, analogue to gravitational time-dilation.
Here we adopt a non-perturbative approach and present a finite-dimensional generalisation of this mechanism, expressing the quantum time-dilation effect as an effective interaction involving previously non-interacting
system components. We name this a \textit{Time-Dilation induced Interaction Transfer (TiDIT) mechanism} and discuss an example using two coupled spins as a quantum clock model. 
Our approach is suitable for implementations in current quantum technology and provides a new tool for exploring gravity at the intersection with quantum physics.
\end{abstract}

\maketitle

%%%%%%%%%%%%%%%%%%%%%%%%%%%%%%%%%%%%%%%%%%%
% DOCUMENT BODY
%%%%%%%%%%%%%%%%%%%%%%%%%%%%%%%%%%%%%%%%%%%

\section{Introduction}\label{sec:Intro}

In general relativity, an ideal clock is assigned to each world line via an underlying metric structure. This notion of time enters the Schrödinger equation as an ideal $t$-parameter and is employed to describe the time evolution of quantum systems moving along any given world line. 
Similarly, an appropriate background structure is required in ordinary Quantum Field Theory in Curved Spacetime, where canonical commutation relations are imposed on equal-time hypersurfaces.
However, assuming the quantum superposition principle and the relativistic mass-energy equivalence to be simultaneously true, matter distributions can be in a superposition of energy and momentum eigenstates. In that case, Einstein's field equations do not suffice to describe a definite space-time structure, hence making ill-defined the notion of an ideal background time parameter~\cite{CastroRuizEtAl20,Isham92}.

%%% Thus causal relationships, and in particular the notion of ‘spacelike’, appear to depend on the quantum state. Does this mean that ‘time’ also is state dependent? Pag.14

In the absence of a general framework for addressing such situations, various approaches have been proposed; yet all encounter significant challenges due to the incompatibility between the roles that time plays in general relativity and quantum mechanics. Collectively, these difficulties are referred to as the so-called \textit{Problem of Time}~\cite{KucharK92,Isham92,Anderson12,BojowaldEtAl11_A,BojowaldEtAl11_B,HoehnEtAl21}, the most common formulation being that, if the Hamiltonian of the system in the classical theory is constrained to vanish, such as in general relativity, then the physical states in the quantum theory do not evolve in time.

To solve the issue, Page and Wootters (PaW) argued that, if the Universe is an isolated quantum system, there is no place in the theory for an external time parameter and time must be measured with a physical clock that is part of the system itself \cite{GiovannettiEtAl15,Wootters84,PageW83,MarlettoV16}.
Hence, the main idea of their proposal, known as the PaW mechanism, is to identify one of the physical degrees of freedom of a quantum system as a clock for the rest of the system, where the flow of time is encoded in the quantum correlations built between the subsystems identified in this way.

The main problems related to this proposal are referred to as Kucha\v{r}'s criticisms \cite{KucharK92}: the fact that i) when applied to a relativistic particle on a Minkowski background, the prediction for its localization probability differs from the accepted Klein-Gordon probability density, 
ii) applying the projection postulate to time measurements violates the Hamiltonian constraint, throwing the original state out of the physical Hilbert state of the system, and
iii) the so-derived conditional probabilities at different times give unphysical results. 
These problems have only recently been addressed and partially solved by the language of quantum information~\cite{GiovannettiEtAl15,HoehnEtAl21}.
In particular, using a POVM to define a \textit{Time operator}~\cite{Pegg98, GiovannettiEtAl15} for an infinite-dimensional quantum clock, i) the derivation of an acceptable localization probability for a relativistic particle is presented in \cite{HoehnEtAl20}, ii) the compatibility of the measurement process with the Hamiltonian constraint is addressed in \cite{HoehnEtAl21}, and iii) the correct statistical description for sequential-time measurements is recovered in \cite{HoehnEtAl21} and \cite{GiovannettiEtAl15} via different approaches, whose comparison is given in \cite{HausmannEtAl23}.
The new setting has been employed to study the time-localizability of events in the presence of gravitational interaction between many quantum clocks \cite{SmithM19, SinghFr23, CastroRuizEtAl17, CastroRuizEtAl20, FavalliS23, DeLaHametteEtAl23}, inspired new approaches in the description of time via generalized coherent states \cite{FotiEtAl21,CoppoEtAl23} and paved the way for new avenues of inquiry in the quantum description of the space-time structure \cite{SmithM20,DiazEtAl19_A,DiazEtAl19_B,DiazEtAl21,Giacomini21,FavalliS22,GiovannettiEtAl23,HoehnEtAl23,DeVuystEtAl24}. 

In this paper, we adopt an operational approach and present a reformulation of the PaW mechanism for finite-dimensional quantum systems that can be easily generalized to different clock models. 
This may be particularly useful for practical applications as, if an implementation of quantum time with quantum technologies is realised, this should involve most likely finite-dimensional clocks. 
This framework is consistent with the results obtained in the ideal clock case - recovered in the infinite-dimensional limit - but also yields a simpler and non-perturbative expression of the Schrödinger equation when the adopted clock is affected by a gravitational-like interaction.
In particular, we consider an isolated quantum system consisting of two non-interacting components, one of which serves as a clock system for the other. We introduce a composite clock model as a network of finite-dimensional quantum systems, which can also be used as individual clocks.
One can think of the composite (global) clock time as labelling space-like hypersurfaces embedding the spatially localized network of (local) clocks \cite{SmithM19}. 
We observe that the \textit{proper time} of a local clock and the corresponding dynamical description differ from the global ones in the presence of a gravitational-like interaction between the individual clocks.
As in the ideal clock case, this interaction gives rise to an effect called a \textit{quantum time-dilation}, in analogy to gravitational time-dilation~\cite{CastroRuizEtAl20}. 
This effect can be interpreted in our framework as an interaction transfer between the adopted local clock and the rest of the system, which we name a \textit{Time-Dilation induced Interaction Transfer (TiDIT) mechanism}.
As a consequence, assuming a spin network as the global clock model, we show that a spin-spin interaction can provide an example of a gravitational-like interaction and a non-perturbative effective Hamiltonian can be derived for the rest of the system.

This general framework for finite-dimensional quantum systems and the TiDIT mechanism are the main results of our work, fostering 
the conceptual design of a suited experimental implementation and possible speculations on the quantum gravity side.

This paper is organized as follows: in Section~\ref{sec:2}, we introduce a finite-dimensional quantum clock and derive the Schrödinger equation for the rest of the system; in Section~\ref{sec:3}, we generalize the previous description to the case of a composite clock model, and in Section~\ref{sec:4} we introduce a gravitational-like interaction between its components, leading to the Time-Dilated Schrödinger equation and the above mentioned TiDIT mechanism;
finally, in Section~\ref{sec:5}, we present an implementation of this framework with a clock model made by two two-level systems and discuss the main implications of the TiDIT mechanism on the dynamical description of the system. 

Discussions on how this framework is related to the problem of time in quantum gravity, its potential implications and perspectives for future works are listed in the final section~\ref{sec:Discussion}.

\section{Page and Wootters mechanism}
\label{sec:2}

\subsection{Defining a Quantum Clock system}\label{sec:q_clock}

We consider a quantum system $C$ with associated Hilbert space $\hilbert_C$ of dimension $d_C$. Given a non-degenerate Hermitian operator $\hamilt_C$, we can choose its eigenstates $\{ \ket{w_k} \}_{k=0, ..., d_C - 1} $ to satisfy the relations 
\begin{equation}
    \hamilt_C \ket{w_k} = \hbar w_k \ket{w_k} ,\quad \braket{w_i}{w_j} = \delta_{ij} \,,
\end{equation}
and hence provide an orthonormal basis of $\hilbert_C$. Furthermore, we can define the unitary transformation
\begin{equation}\label{eq:V_C}
    \hat{V}_C(t)=e^{-\frac{i}{\hbar} \hamilt_C t}~,
\end{equation}
associated to the operator $\hamilt_C$ and parametrized by the real positive parameter $t$. Starting from a generic reference state $\ket{\phi} \in \hilbert_C$, we use the operator \eqref{eq:V_C} to define the set of states
\begin{equation}
    \ket{\phi(t)} = \hat{V}_C(t) \ket{\phi} = e^{-\frac{i}{\hbar} \hamilt_C t} \ket{\phi} \,.
\end{equation}
We refer to the system $C$ as a \textit{clock} system and, when the chosen operator $\hamilt_C$  corresponds to the system Hamiltonian, to $\hat{V}_C(t)$ as the \textit{clock evolution operator}. 

Since the action of the clock evolution operator on the eigenstates of the clock Hamiltonian is trivial, a good definition of reference state should have a homogeneous overlap over the whole eigenspectrum. Thus, we define the \textit{clock reference state} $\ket{R}$ as
\begin{equation}\label{eq:referencestate}
    \ket{R}:= \frac{1}{\sqrt{d_C}} \sum_{k=0}^{d_C - 1} e^{-i\varphi_k} \ket{w_k} \,,
\end{equation}
where $\varphi_k$ are arbitrary real coefficients~\cite{WoodsEtAl18,HoehnEtAl21}. The action of the evolution operator on the reference state assigns each superposition component a different time-dependent phase, allowing us to define a \textit{time state}
\begin{equation}\label{eq:timestate}
    \ket{t} := \hat{V}_C(t) \ket{R} = \frac{1}{\sqrt{d_C}}\sum_{k=0}^{d_C - 1} e^{-i ( w_k t + \varphi_k)} \ket{w_k} \,;
\end{equation}
by this construction, we can state that $t$ plays the role of the \textit{proper time} of the clock. In relativity, it can be identified with the time elapsed along the clock's word line \cite{CastroRuizEtAl17,CastroRuizEtAl20}. More generally, we refer to the proper time as an intrinsic property of the quantum clock, i.e. a way to parameterize its time states.

Before moving on, we briefly introduce two fundamental properties of time states which characterize quantum clocks and, in particular, the role of their proper time in describing the dynamics of another quantum system. The first is a direct consequence of the evolution operator and is given by
\begin{equation}\label{eq:property}
    \hamilt_C \, \hat{V}_C(t) 
    = i \hbar  \dv{t} \, \hat{V}_C(t) \,.
\end{equation}
The second, derived in Appendix~\ref{app:Res}, consists of a general expression of the resolution of the Identity
\begin{equation}\label{eq:resolution}
    \iden_C = \int \dd \mu(t) \ketbra{t} \,,
\end{equation}
where the integration measure $\dd \mu(t)$ represents either a discrete sum or a continuous integration. As we shall see, the results discussed in the rest of the article are independent of the particular expression of Eq.~\eqref{eq:resolution}, which however characterizes how the notion of quantum history is encoded in quantum correlations.

\subsection{Evolution without Evolution}

\begin{figure}
    \centering
    \includegraphics[width=0.45\textwidth]{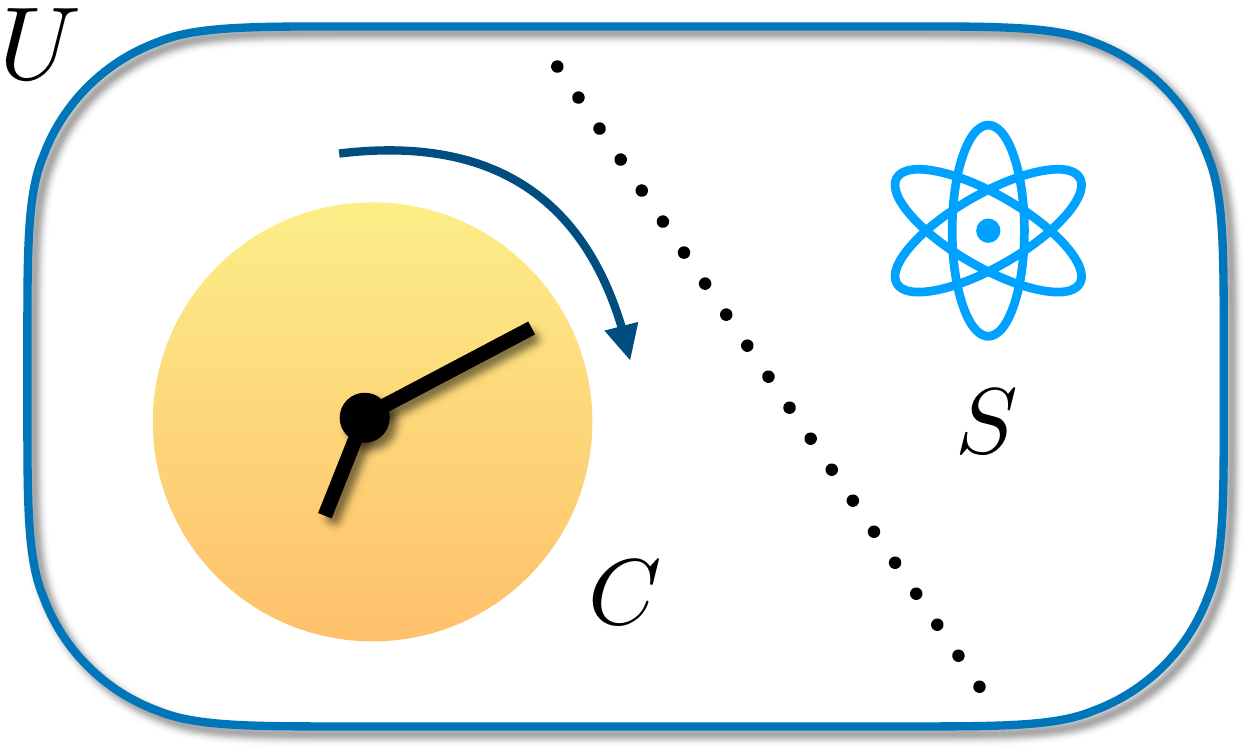}
    \caption{Schematic representation of the Universe system $U$. This is an isolated and bipartite quantum system whose components $C$ and $S$ are assumed to be non-interacting.}
    \label{fig:univ}
\end{figure}

We now consider an isolated bipartite quantum system called the Universe $U$, composed of the clock system $C$, defined in Section~\ref{sec:q_clock}, and a generic quantum system $S$. Therefore, we describe $U$ by a Hilbert space $\hilbert = \hilbert_C \otimes \hilbert_S$ with dimension $d = d_C \cdot d_S$. This configuration is represented in Figure \ref{fig:univ}.
Given the Hamiltonian of the Universe $\hamilt$, any state $\ket{\Psi} \in \hilbert$ satisfying the constraint equation 
\begin{equation}\label{eq:wheeler-dewitt}
    \hamilt \ket{\Psi} = 0 \,,
\end{equation}
is a stationary state with respect to a hypothetical observer that is external to the Universe\footnote{For example, Eq.~\eqref{eq:wheeler-dewitt} describes a stationary Schrödinger equation for an isolated quantum system in a laboratory.}. We can express $\ket{\Psi}$ by the Schmidt decomposition
\begin{equation}\label{eq:schmidt}
    \ket{\Psi} = \sum_{n=0}^{r-1} \sqrt{\lambda_n} \ket{\phi_n}_C\ket{\xi_n}_S \,,
\end{equation}
where the sets $\{\ket{\phi_n}_C\} $ and $\{\ket{\xi_n}_S\}$ are orthonormal and $r \leq min\{ d_C, d_S \}$ is called the Schmidt rank. For any entangled state, we have $r > 1 $. Inserting the resolution of the Identity~\eqref{eq:resolution} in Eq.~\eqref{eq:schmidt}, we get
\begin{equation}\label{eq:history_state}
    \ket{\Psi} =
    \int \dd \mu(t) \; a(t) \ket{t}_C \ket{\psi(t)}_S \,,
\end{equation}
in which
\begin{equation}\label{eq:state_Schr_cont_1}
    a(t) = \norm{\bra{t}_C\ket{\Psi}}\,,
\end{equation}
and
\begin{equation}\label{eq:state_Schr_cont_2}
    \ket{\psi(t)}_S =  \frac{1}{a(t)} \bra{t}_C\ket{\Psi}
\end{equation}
is well defined for all $t$ s.t. $a(t)\neq 0$. When $\ket{\Psi}$ is given in the form of equation \eqref{eq:history_state} we call it the \textit{history state} of the system, and read it as a superposition of states whose components are the product of time states and the so-called \textit{conditional Schrödinger states}  $\ket{\psi(t)}_S$ for the $S$ system \cite{PageW83,GiovannettiEtAl15}. 

Using Eq.~\eqref{eq:history_state} one can show that the dynamics of $S$ can be described in terms of the proper time of the subsystem $C$; in this case, the time evolution - and thus the quantum history - of $S$ is encoded in its quantum correlations with $C$, and independent of any external laboratory time. 
Indeed, the constraint equation \eqref{eq:wheeler-dewitt} ensures not only that the history state is globally stationary, but also allows the recovery of the Schrödinger equation for one subsystem using the other as a clock.
To see this, we consider the non-interacting Hamiltonian
\begin{equation}\label{eq:Hamiltonian_CG}
    \hamilt = \hamilt_C \otimes \iden_S + \iden_C \otimes \hamilt_S \,,
\end{equation}
and condition the constraint equation~\eqref{eq:wheeler-dewitt} on a time state $\ket{t}_C$, obtaining
\begin{equation}
    \bra{t}_C \hamilt \ket{\Psi} 
    = \bra{t}_C \hamilt_C \ket{\Psi} + \bra{t}_C \hamilt_S \ket{\Psi} = 0\,.
\end{equation}
Recalling the property \eqref{eq:property} and the definition of a conditional Schrödinger state \eqref{eq:state_Schr_cont_2} we obtain
\begin{equation}\label{eq:Schr_chi}
    i \hbar \dv{t} \ket{\psi(t)}_S =  \hamilt_S \ket{\psi(t)}_S \,,
\end{equation}
which is the conventional Schrödinger equation for the system $S$ with respect to $C$'s proper time.
As discussed in Appendix~\ref{app:lambda}, the non-interacting Hamiltonian~\eqref{eq:Hamiltonian_CG} implies that the norm of the conditional state does not vary in time, i.e. $a(t) = a(0) = const$. A contribution to the dynamical equations can emerge from the normalization coefficient $a(t)$ when considering a different constraint operator.

An equivalent description of the dynamics is provided by the Liouville-von Neumann equation. In the density matrix formalism, the conditional state for the system $S$ is given by
\begin{multline}\label{eq:conditionalmatrixglobal}
    \rho_S (t) 
    := \frac{\Tr_{C}\left[\ketbra{t}_C \,\rho \,\right]}{\Tr\left[\ketbra{t}_C \,\rho \,\right]} 
    = \frac{\expval{\rho}{t}_C}{a^2 (t)} =
    \\
    = \frac{\expval{\ketbra{\Psi}}{t}_C}{a^2 (t)} 
    = \ketbra{\psi(t)}_S\,,
\end{multline}
and the corresponding von Neumann equation is
\begin{equation}\label{eq:composite_dense_global}
    \dv{t} \rho_S (t) 
    =
    - \frac{i}{\hbar} \left[ \hamilt_S , \rho_S (t) \right]
    \,,
\end{equation}
in which again $a(t) = const$.
The state $\rho_S(t)$ is also called an Everett relative state \cite{MarlettoV16,Everett57}.

This expression can be employed to evaluate $S$'s observables and thus study their dynamical evolution. We define the expectation value of the observable $\observable_S$ at the \textit{internal} time $t$ as
\begin{equation}\label{eq:averageglobal}
    \expval{\observable_S} (t) 
    := \frac{\Tr[ \observable_S \, \ketbra{t}_C \,  \rho \,]}{\Tr\big[\ketbra{t}_C \, \rho \,\big]} 
    = \Tr_S[\observable_S \, \rho_S (t) \, ] \,.
\end{equation}
Its time evolution thus satisfies the relation
\begin{equation}\label{eq:averageglobalintime}
    \dv{t} \expval{\observable_S} (t) = -\frac{i}{\hbar} \Tr_S\left[\left[\observable_S \,, \hamilt_S \right] \rho_S(t) \, \right] \,,
\end{equation}
from which the conventional Heisenberg equation for the operator $\observable_S (t)$ is easily recovered.

In the literature, the choice of $a(t)$ remains in general ambiguous \cite{GiovannettiEtAl15} and the condition $a(t) = constant$ is naturally implemented starting from a homogeneous history state. Here instead, starting from the Hamiltonian of the finite-dimensional Universe, we give a prescription on how to derive the expression of the history state from a generic solution of the constraint equation and uniquely fix this coefficient.
The dynamical laws of quantum mechanics can thus emerge as a phenomenological consequence of a relational description of time evolution. 
This picture of the quantum dynamics as a description of the quantum correlations between the clock system and the rest of the Universe is called the Page and Wootters (PaW) mechanism or, as its original proponents poetically said in their work, \textit{Evolution without Evolution} \cite{PageW83}.

\section{Extension to a composite clock}
\label{sec:3}

\subsection{Local clock systems}

We now consider the Universe system $U$ introduced in Section~\ref{sec:2} and focus on the internal structure of the clock system $C$. In particular, we assume it is composed of $N$ non-interacting sub-clocks, labelled by an index $J = A, B, ..., N$, each associated with a Hilbert space $\hilbert_J$ of dimension $d_J$ and a non-degenerate Hamiltonian $\hamilt_J$. Even if we do not assume any spatial structure or ordering within $C$, we will often refer to these sub-clocks as local clock systems, and to $C$ as the global clock. The dimension of the global clock's Hilbert space $\hilbert_C = \otimes_J \hilbert_J$ is given by $d_C = \prod_J d_J $ and its Hamiltonian can be expressed as
\begin{equation}\label{eq:global_clock}
    \hamilt_C = \sum_J \hamilt_J \,.
\end{equation}
It is crucial to notice that this construction implies the local clocks are non-interacting. This assumption helps us understand how the dynamical evolution can be described when adopting a local clock, and we will relax it by taking into account a specific family of interaction terms in Section~\ref{sec:4}.

Following the same steps of Section~\ref{sec:2}, we introduce an energy eigenbasis $\{ \ket{w^{J}_k}_J \}_{k = 0, ..., d_J} \,,$ and an evolution operator $\hat{V}_J(\tau):= \exp\{-\frac{i}{\hbar} \hamilt_J \tau\}$ for each local clock $J$. 
We can thus introduce the notion of a local time state via the relation $\ket{\tau}_J:= \hat{V}_J(\tau) \ket{R}_J$, where the local reference state $\ket{R}_J$ is defined in analogy with Eq.~\eqref{eq:referencestate}, and the parameter $\tau$ represents the proper time of a local clock.
The relation
\begin{equation}\label{eq:mappingevolution}
    \hat{V}_C(t) = e^{-\frac{i}{\hbar} \hamilt_C t} 
    = \otimes_J \, e^{-\frac{i}{\hbar} \hamilt_J t} = \otimes_J \hat{V}_J (t)
\end{equation}
between the global evolution operator $\hat{V}_C(t)$ and the local operators $\hat{V}_J(t)$ implies that the global time states can be written in terms of the local ones as
\begin{equation}\label{eq:referencestate_corresp}
    \ket{t}_C = \otimes_J \ket{t}_J \,
    \text{, with } \, 
    \ket{R}_C := \otimes_J \ket{R}_J \,.
\end{equation}
One can also show that choosing the coefficients of the local reference states automatically fixes the global ones. Finally, an expression for the resolution of the Identity in terms of the local time states is given by
\begin{equation}\label{eq:resolution_sub}
    \iden_C = \otimes_J \iden_J 
    = \otimes_J \int \dd \mu_J(\tau_J)  \ketbra{\tau_{J}}_J \,,
\end{equation}
where we introduced a set of local time parameters $\{\tau_J\}_{J = A, ..., N} $, each associated with a local time state of the corresponding $J$-th sub-clock.

\subsection{History from local clocks}

The quantum history of $U$'s components can be described from the perspective of a local clock, i.e. using its proper time to describe time evolution. To see this, we start by making explicit the relation between the global clock state and the local ones in the history state. This is obtained from Eq.~\eqref{eq:history_state} by inserting the resolution of the Identity \eqref{eq:resolution_sub} as
\begin{multline}\label{eq:historystatemultiple}
    \ket{\Psi} = \int  \dd \mu(t) \dd \mu_A(\tau_A) \dots \dd \mu_N(\tau_N) \times
    \\ 
    \times a(t) \, F(\{\tau_{J}\}| \, t \, ) \big(\otimes_J \ket{\tau_{J}}_J \big) \ket{\psi (t)}_S \,,
\end{multline}
in which $F(\{\tau_J\} | \,t\, )$ is given by
\begin{equation}\label{eq:transition_subclocks}
    F(\{\tau_J\} | \,t\, ) := \bigg( \otimes_J \bra{\tau_J}_J \bigg) \ket{t}_C = \prod_J \bra{\tau_J}\ket{t}_J \,,
\end{equation}
and describes the transition amplitude between the global time state $ \ket{t}_C$ and the global clock state associated with a given set of local time parameters $\{\tau_J\}$. 
As discussed in Appendix \ref{app:Res}, different time states are generally non-orthogonal and thus the transition amplitudes $\bra{\tau_J}\ket{\,t\,}_J$ are generally non-zero.

This finite-dimensional framework can be easily generalized to many different clock models, and the ideal clock description can be easily recovered by taking the infinite-dimensional limit for each $J$, where these amplitudes become delta functions. In this case, the history state~\eqref{eq:historystatemultiple} can be expressed as
\begin{equation}\label{eq:infinite}
    \ket{\Psi} = \frac{1}{T} \int_{-T/2}^{+T/2} \dd t \, a(t) \big(  \otimes_J^N \ket{\bar{t} \,}_J \big) \ket{\psi (t)}_S \,,
\end{equation}
where the non-normalized states $\ket{\bar{t} \,}_J := \sqrt{d_J} \ket{\, t \,}_J$ are now orthogonal in $[-T/2, T/2)$ with period $T$. Moreover, the state~\eqref{eq:infinite} with $a(t) = 1$ reduces to the expression commonly used in the literature \cite{GiovannettiEtAl15,CastroRuizEtAl20,SmithM19} by taking the continuous spectrum limit, i.e. $T \to \infty$. One can also show that the description of bounded and continuous spectrum clock models~\cite{HausmannEtAl23} is recovered via a similar limit procedure.

From the perspective of the sub-clock $A$, i.e. in terms of its proper time $\tau \equiv \tau_A$, the expression of the history state~\eqref{eq:historystatemultiple} becomes
\begin{equation}\label{eq:historyfromsubclock}
    \ket{\Psi} = \int  \dd \mu_A(\tau) \, a_A(\tau) \ket{\tau}_A \ket{\psi(\tau)}_{U|A} 
     \,,
\end{equation}
in which, for every $a_A(\tau) = \norm{\bra{\tau}_A\ket{\Psi}} \neq 0$, we have
\begin{equation}\label{eq:conditional_Sch}
    \ket{\psi(\tau)}_{U|A} := \frac{1}{a_A(\tau)} \bra{\tau}_A\ket{\Psi}
    \,,
\end{equation}
where $\ket{\psi(\tau)}_{U|A}$ is a conditional Schrödinger state for the rest of the Universe normalized by $a_A(\tau)$. As mentioned in the previous section, this normalization coefficient must be constant in $\tau$ when the adopted clock does not interact with the rest of the Universe. 

\subsection{Dynamical evolution}

To recover the dynamical equations from the perspective of a local clock, we follow the same procedure described in Section~\ref{sec:2}.
Conditioning the non-interacting constraint~\eqref{eq:Hamiltonian_CG} now on the time state $\ket{\tau}_A$ of the local clock $A$, one gets
\begin{equation}\label{eq:composite_Schro}
    \bra{\tau}_A \bigg( \hamilt_S + \hamilt_A + \sum_{J \neq A} \hamilt_J \bigg) \ket{\Psi} = 0
\end{equation}
and, recalling the property in Eq.~\eqref{eq:property}, we obtain
\begin{equation}\label{eq:Schr_multipl_chi}
    i \hbar {\dv{\tau}} \ket{\psi(\tau)}_{U|A} 
    =  \hamilt^{{(A)}} \ket{\psi(\tau)}_{U|A}\,,
    \end{equation}
where
\begin{equation}\label{eq:HamiltonianEffectiveNonInteract}
    \hamilt^{{(A)}} := \hamilt_S + \sum_{J \neq A} \hamilt_J \,.
\end{equation}
This relation describes the dynamical evolution of the conditional Schrödinger state $\ket{\psi(\tau)}_{U|A}$ for the rest of the Universe in terms of $A$'s proper time.

Finally, the dynamical evolution of a generic observable acting on $U|A$ can be described in the density matrix formalism introducing the conditional state for the rest system in analogy with Eq.~\eqref{eq:conditionalmatrixglobal} as
\begin{equation}\label{eq:conditionalmatrix}
    \rho^{(A)} (\tau) 
    = \frac{\Tr_{A}\left[\ketbra{\tau}_A \,\rho \,\right]}{\Tr\left[\ketbra{\tau}_A \,\rho \,\right]} 
    = \frac{\expval{\ketbra{\Psi}}{\tau}_A}{a^2_A(\tau)} \,.
\end{equation}
The von Neumann equation is then given by
\begin{equation}\label{eq:composite_dense}
    \dv{\tau} \rho^{(A)} (\tau)
    =
    - \frac{i}{\hbar} \left[ \, \hamilt^{(A)} \, , \, \rho^{(A)}(\tau) \, \right]
    \,,
\end{equation}
while the evolution of the expectation value of the operator $\observable_S$ in terms of $A$'s proper time, defined as
\begin{equation}
    \expval{\observable_S^{(A)}} (\tau) 
    = \Tr_{U|A}[\, \observable_S \, \rho^{(A)} (\tau) \, ]
\end{equation}
in analogy with Eq. \eqref{eq:averageglobal}, satisfies the relation
\begin{multline}\label{eq:observablesevolution}
    \dv{\tau} \expval{\observable_S^{(A)}} = \Tr_{U|A}\bigg[\, \observable_S \, \dv{\tau} \, \rho^{(A)} (\tau)\, \bigg] =
    \\
    = -\frac{i}{\hbar} \Tr_{U|A}\left[ \left[ \, \observable_S \,, \, \hamilt^{(A)} \, \right] 
     \rho^{(A)} (\tau)\, \right] \,.
\end{multline}
Comparing this result with Eq.~\eqref{eq:averageglobalintime}, we observe that the dynamical descriptions observed by different clocks are equivalent in the absence of interaction. 

\section{TiDiT mechanism}\label{sec:4}

Interactions between the components of the clock partition $C$ do not affect the dynamics from the global clock perspective: the dynamical laws derived in Section~\ref{sec:2} only rely on the absence of interaction between the system adopted as a clock and the rest of the Universe $U$. Hence, the question naturally emerges: \textit{what changes in the local description of time} when interactions among the global clock's components are taken into account? 
In this section, we show that allowing one particular kind of interaction between the local clocks only affects their proper time as seen from their perspective. The resulting description is particularly tractable when choosing the interaction to be gravitational-like, i.e. a pairwise interaction given by the tensor product of the local free Hamiltonians. While somewhat special, this interaction is not new \cite{CastroRuizEtAl20, SmithM19} and allows discussing a quantum model for time dilation, which is one of the main results of this article.

\subsection{Gravitational-like Interaction}

Introducing a generic interaction between the local clocks prevents a time-local description of dynamical evolution from their perspective \cite{SmithM19}. However, when the interaction Hamiltonian $\hamilt_{int}$ commutes with the local Hamiltonians $\hamilt_J$ of the adopted clock, we can still use the machinery of the previous sections to derive the dynamical equations. In particular, let us consider an interaction term as a sum of contributions each proportional to the tensor product of pairs of local clocks free Hamiltonians, i.e. a \textit{gravitational-like interaction} given the analogy with the Newtonian potential \cite{CastroRuizEtAl17,CastroRuizEtAl20}. By this choice, the global clock Hamiltonian \eqref{eq:global_clock} becomes
\begin{equation}\label{eq:gravitational-hamiltonian}
    \hamilt_C = \sum_J \hamilt_J + \hamilt_{int} \,,
\end{equation}
where
\begin{equation}\label{eq:gravitational-likeinteraction}
        \hamilt_{int} = - \frac{1}{2} \sum_{J, K} g_{JK} \, \hamilt_J \hamilt_K \,,   
\end{equation}
and the $g_{JK}$ are generic coupling constants such that for each local clock $J$ we have $g_{JJ} = 0$.

Introducing this interaction term, the relations between the global and local evolution operators and time states become
\begin{equation}
    \hat{V}_C(t) = e^{-\frac{i}{\hbar} \hamilt_{int}t} \left( \otimes_J^N \hat{V}_J (t) \right) \,,
\end{equation}
and
\begin{equation}
    \ket{t}_C = e^{-\frac{i}{\hbar} \hamilt_{int}t} \otimes_J^N \ket{t}_J \,.
\end{equation}
As a consequence, the transition amplitude between different time states
\begin{equation}\label{eq:amplitude_composite}
    F( \{\tau_J\} | t ) = \bigg( \otimes_J^N \bra{\tau_J}_J \bigg) e^{- \frac{i}{\hbar} \hamilt_{int}t} \bigg( \otimes_J^N \ket{t}_J \bigg)
\end{equation}
cannot be expressed as a product of contributions from each local clock as in Eq.\,\eqref{eq:transition_subclocks}. It is important to note that, when expressing the history state in terms of the proper time of a clock, the general structure in Eq.~\eqref{eq:historystatemultiple} and Eq.~\eqref{eq:historyfromsubclock} remains unchanged. Also, the definitions of the conditional Schrödinger state \eqref{eq:conditional_Sch} and conditional density matrix \eqref{eq:conditionalmatrix} still hold in the presence of a gravitational-like interaction.

\subsection{Time-Dilated Schrödinger equation}

In the presence of gravitational-like interaction, a generalization of the Schrödinger equation can be derived following the same steps of the previous sections and making use of the relation
\begin{equation}
    \bra{\tau}_A g_{AJ} \hamilt_A \hamilt_J \ket{\Psi} 
    = - i \hbar \, g_{AJ} \hamilt_J \dv{\tau}\bra{\tau}_A \ket{\Psi}\,,
\end{equation}
which is a generalization of the property in Eq.~\eqref{eq:property}. The result is a \textit{Time-Dilated Schrödinger equation} which, from the perspective of the clock $A$, reads
\begin{equation}\label{eq:SchrInteractGeneral}
    i \hbar \hat{R}(A) \dv{\tau} \ket{\psi(\tau)}_{U|A} 
    =\hamilt^{(A)} \ket{\psi(\tau)}_{U|A}\,,
\end{equation}
where
\begin{equation}
    \hamilt^{(A)} := 
    \hamilt_S 
    + \sum_{J\neq A} \hamilt_J
    - \frac{1}{2}\sum_{J, K \neq A} g_{JK} \, \hamilt_J \hamilt_K \,,
\end{equation}
is the conditional Hamiltonian introduced of the non-interacting case~\eqref{eq:HamiltonianEffectiveNonInteract} with the addition of the gravitational-like interaction between the other local clocks, and where the operator $\hat{R}(A)$, defined as
\begin{equation}\label{eq:redshiftopgeneral}
    \hat{R}(A) := \left( \iden - \sum_{J \neq A} g_{AJ} \hamilt_J \right) \,,
\end{equation}
can be read as a \textit{redshift operator} which leads to a \textit{quantum time-dilation} effect. Its action accounts for the interaction of the adopted clock $A$ with the rest of the Universe, and provided that the relative evolution is still unitary, we again have $a_A(\tau) = const $.

This operator was first introduced in Ref.~\cite{CastroRuizEtAl20} in the context of \textit{Time Reference Frames}. There, an analogue result to Eq.~\eqref{eq:SchrInteractGeneral} has been derived while discussing temporal localization of events with respect to gravitating quantum clocks. The authors also showed that $\hat{R}$ plays the role of the $g_{00}$ component of the Schwarzschild metric in the weak field approximation, describing a \textit{gravitational time-dilation} effect. In the following, we focus on the consequences of this dynamical description in the finite-dimensional case, leaving the implications of exploring gravity at the intersection with quantum mechanics to future works.

\subsection{The TiDIT mechanism}

The first consequence of this finite-dimensional framework is the possibility to rigorously cast the Time-Dilated Schrödinger equation~\eqref{eq:SchrInteractGeneral} into a simpler form. This can be obtained non-perturbatively when the redshift operator $\hat{R}(A)$ is invertible. To show this, we first define the operator
\begin{equation}\label{eq:ratioOp}
     \hat{\Phi}(A) \equiv
     \sum_{J\neq A} g_{AJ} \hamilt_J \,,
\end{equation}
and hence rewrite Eq.~\eqref{eq:redshiftopgeneral} as
\begin{equation}
    \hat{R}(A) = \left( \iden - \hat{\Phi}(A)\right) \,.
\end{equation}
If the spectral radius\footnote{The spectral radius $\rho(\hat{O})$ of an operator $\hat{O}:\hilbert\rightarrow\hilbert$ from a finite-dimensional Hilbert space to itself is defined as the maximum of the absolute values of its eigenvalues.} of the operator $\hat{\Phi}(A)$ is smaller than one, i.e. $\rho(\hat{\Phi}(A)) < 1$, the inverse of the redshift operator is given via the geometric series as
\begin{equation}\label{eq:geometric}
    \hat{R}^{-1}(A) = \left( \iden - \hat{\Phi}(A) \right)^{-1} = \sum_{n=0}^{+\infty} \hat{\Phi}^n(A) \,.
\end{equation}
Multiplying the generalized Schrödinger equation~\eqref{eq:SchrInteractGeneral} by $\hat{R}^{-1}(A)$ from the left, we obtain
\begin{equation}
    i \hbar \dv{\tau} \ket{\psi(\tau)}_{U|A}
    = \, \sum_{n=0}^{+\infty} \hat{\Phi}^n(A) \, \hamilt^{(A)}
    \ket{\psi(\tau)}_{U|A}\,.
\end{equation}
This is the dynamical equation obtained in the non-interacting case plus contributions entering in powers of the interaction couplings $g_{JK}$. Indeed, expanding the expression above and substituting the explicit expression for $\hat{\Phi}(A)$ from Eq.~\eqref{eq:ratioOp}, we obtain
\begin{widetext}
\begin{multline}\label{eq:TiDITgravity}
    \hamilt^{(A)}_{eff} := \sum_{n=0}^{+\infty} \hat{\Phi}^n(A)\, \hamilt^{(A)} = 
    \hamilt_S 
    + \sum_{J \neq A} \hamilt_J
    - \frac{1}{2} \sum_{J, K \neq A} \left( g_{JK} - 2g_{AJ}\right) \, \hamilt_J \,\hamilt_K
    \, + \\
    + \sum_{J \neq A} \, g_{AJ} \, \hamilt_J \, \hamilt_S 
    + \sum_{J,K \neq A} g_{AJ} \, g_{AK} \,  \hamilt_J \, \hamilt_K \, \hamilt_S
    -  \frac{1}{2} \sum_{J, K, L \neq A} g_{AL} \left( g_{KJ} - 2g_{AJ}\right) \, \hamilt_J \, \hamilt_K \, \hamilt_L + O(g_{JK}^3)
    \,.
\end{multline}
\end{widetext}
In this expression, the gravitational-like interactions involving the clock $A$ have been transferred to the system $S$ and the other local clocks, giving rise to a re-scaling of the $2$-body couplings $g_{JK}$ and new $n$-body interactions. Moreover, the system $S$, which represents the globally non-interacting degrees of freedom of the Universe, is coupled with the other local clocks from $A$’s perspective. 
These couplings derive from an equivalent expression of the Time-Dilated Schrödinger equation \eqref{eq:SchrInteractGeneral} and can thus be read as the translation of the quantum time-dilation effect in terms of new effective interactions. We call this a \textit{Time-Dilation induced Interaction Transfer (TiDIT) mechanism}. 

Furthermore, the TiDIT mechanism can be employed to exactly derive the dynamical evolution of a generic observable $\observable$ acting on $U|A$. Indeed, the generalization of Eq.~\eqref{eq:observablesevolution} to the case of gravitational-like interacting quantum clocks gives
\begin{multline}\label{eq:redshift_observ_dynamics}
    \dv{\tau} \expval{\observable^{(A)}} (\tau) 
    = \Tr_{U|A}\left[ \, \observable \, \dv{\tau} \, \rho^{(A)} (\tau) \, \right] =
    \\
    = - \frac{i}{\hbar} \Tr_{U|A}\left[ \left[ \, \observable \,, \, \hamilt^{(A)}_{eff} \, \right] \rho^{(A)} (\tau) \, \right]\,.
\end{multline}
As an example, one can easily compute the expectation value of the redshift operator $\hat{R}(A)$ obtaining 
\begin{equation}\label{eq:redshift_dynamics}
    \dv{\tau} \expval{\hat{R}(A)} = 0 \,.
\end{equation}

In the next section, we address the implications of this result by means of examples with spin systems, potentially implementable in current quantum technology platforms.

\section{Example with spin clocks}
\label{sec:5}

\subsection{Two-level clock model}

In order to provide an explicit example that is suitable also for experimental applications, let us consider a global clock system $C$ made by a spin-$1/2$ particle, associated with a Hilbert space of dimension $d_C = 2$ and having Hamiltonian operator
\begin{equation}\label{eq:qubit}
    \hamilt_C = \hbar w \, \hat{\sigma}_C^x \,.
\end{equation}
As we will now demonstrate, this choice enables a straightforward interpretation of the usual quantum computational basis element $\ket{0}$ as a convenient reference state for the preceding construction. Indeed, the Hamiltonian's eigenvalues and eigenvectors can be expressed as $\omega_\pm = \pm\hbar \omega $ and $\ket{\pm}$ respectively\footnote{According to the notation in Section~\ref{sec:2}, the eigenvalues of $\hamilt_C$ can also be denoted by $w_k =  w \, (2k - 1) $ with $k = 0, 1 $.}, and the clock reference state \eqref{eq:referencestate} as
\begin{equation}\label{eq:qubitreference}
    \ket{R}_C = \frac{1}{\sqrt{2}} (\ket{+} + \ket{-}) \,.
\end{equation}
Then, time states are obtained by the map 
\begin{equation}\label{eq:qubitevolut}
    \hat{V}_C(t)=e^{-iwt\hat{\sigma}^x_C}
\end{equation}
as
\begin{equation}\label{eq:qubittimestate}
    \ket{t}_C=\hat{V}_C(t)\ket{R}_C \,.
\end{equation}
Hence, we can rewrite the reference state as the $t_0=0$ time state
\begin{equation}
    \ket{t_0}_C=\hat{V}_C(0)\ket{R}_C=\frac{1}{\sqrt{2}} (\ket{+} + \ket{-})
\end{equation}
and, choosing $t_1=\pi/2w$, the smallest time state orthogonal to the reference as
\begin{equation}\label{eq:qubit_orthogonalization}
     \ket{t_1}_C := \hat{V}_C(\pi/2w)\ket{R}_C = \frac{-i}{\sqrt{2}} (\ket{+} - \ket{-}) \,.
\end{equation}
As it is clear, the set $\{ \ket{t_0}_C, \ket{t_1}_C\}$ provides an orthonormal basis for the clock Hilbert space\footnote{In our notation, $\ket{t_0}_C=\ket{0}_C$ is the lowest eigenstate of the Pauli $\hat{\sigma}^z$ operator and, in the notation of quantum information theory~\cite{NielsenC10}, it  corresponds to the state $\ket{0}$; similarly, $\ket{t_1}_C=\ket{\pi/2w}_C$ corresponds to the state $\ket{1}$, but it is important to stress that $\ket{1}_C\neq\ket{1}$ unless $w = \pi/2$.}, and it is easy to show that
\begin{equation}
    \frac{2w}{\pi} \int_{0}^{\frac{\pi}{w}} \dd t \, \ketbra{t}_C = \iden_C \,.
\end{equation}
Therefore, one can recover a discrete and a continuous resolution of the identity in terms of time states. Inserting them into the history state~\eqref{eq:history_state} we obtain
\begin{equation}\label{eq:history_qubit}
    \ket{\Psi} = a_0 \ket{t_0}_C \ket{\psi(t_0)}_{S} + a_1 \ket{t_1}_C \ket{\psi(t_1)}_{S} \,,
\end{equation}
and 
\begin{equation}\label{eq:history_qubit_cont}
    \ket{\Psi} = \frac{2 w}{\pi} \int_{0}^{\pi/w} \dd t\, a(t) \, \ket{\,t\,}_C \ket{\psi(t)}_{S} \,,
\end{equation}
where
\begin{equation}
    a(t) = \sqrt{a_0^2 \cos^2(wt)  + a_1^2 \sin^2(wt)}\,.
\end{equation}
We highlight that the non-interacting constraint~\eqref{eq:Hamiltonian_CG} again implies $a_1 = a_0 = 1/\sqrt{2}$ and $a(t) = 1/\sqrt{2}$. The expression in Eq.~\eqref{eq:history_qubit} describes two independent time frames of the system's history, while the one in Eq.~\eqref{eq:history_qubit_cont} is a non-orthogonal superposition passing through all intermediate times.

\subsection{$N$-spins clock model}

\begin{figure}
    \centering
    \includegraphics[width=0.45\textwidth]{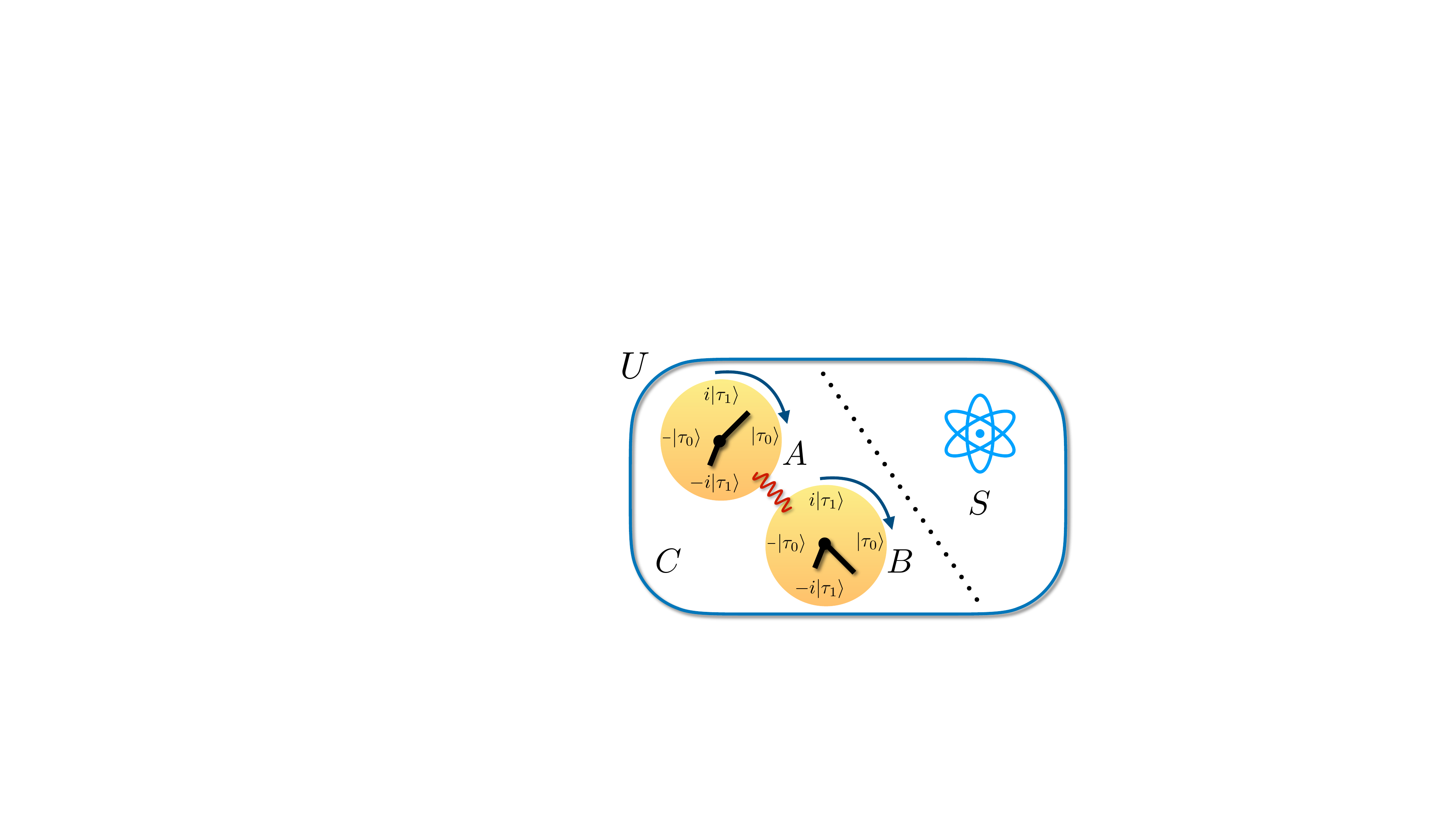}
    \caption{Schematic representation of the Universe system $U$ in Figure \ref{fig:univ} when the global clock $C$ is a composite quantum system. In particular, its components are two eventually interacting quantum spins labelled with $A$ and $B$. In the presence of a spin-spin interaction, they can still be used as individual clock systems to describe the time evolution of the rest of the Universe.}
    \label{fig:spins}
\end{figure}

Let us now consider the case of a global clock system whose components, labelled by $J = A, B,..., N,$ are two-level systems. This configuration is represented in Figure \ref{fig:spins}. Given a Hamiltonian $\hamilt_J$ for each local clock in the form of Eq.~\eqref{eq:qubit}, the global clock Hamiltonian~\eqref{eq:global_clock} for the case $N = 2$ becomes 
\begin{equation}\label{eq:globalclockhamilt}
    \hamilt_C = \sum_J \hamilt_J 
    = \hbar \omega \, \hat{\sigma}_A^x + \alpha \hbar \omega \, \hat{\sigma}_B^x \, ,
\end{equation}
where $\omega=\omega_A$ and the constant
\begin{equation}\label{eq:chain_ref}
    \alpha=\frac{\omega_B}{\omega_A}
\end{equation}
ensures $\hamilt_C$ is non-degenerate for $\lvert\alpha\rvert\neq 1$. 
Given the eigenstates of the local clock Hamiltonians as $\ket{\pm}_J$, we can introduce the local reference states $\ket{R}_J$, evolution operators $\hat{V}_J(\tau)$ and time states $\ket{\tau}_J$ in analogy with Eq.s~(\ref{eq:qubitreference}-\ref{eq:qubit_orthogonalization}). 
Consequently, the eigenstates of the global clock Hamiltonian are given by the four possible tensor products $\ket{\pm}_A\ket{\pm}_B $, the global reference state can be written as
\begin{equation}
    \ket{R}_C = \ket{R}_A \ket{R}_B =
    \frac{1}{2}\sum_{a=\pm} 
    \ket{a}_A \sum_{b=\pm} \ket{b}_B \,,
\end{equation}
and the global time state as 
\begin{equation}
\begin{split}
    \ket{\,t\,}_C 
    &=\hat{V}_C(t)\ket{R}_C = \ket{\,t\,}_A \ket{\,t\,}_B
    \\
    &=\frac{1}{2}\sum_{a,b=\pm}e^{-i\omega(a+\alpha b) t}\ket{a}_A\ket{b}_B \,,
\end{split}
\end{equation}
where $\hat{V}_C(t) = \hat{V}_A(t) \otimes \hat{V}_B(t)$. Whether applying the global evolution operator onto the global reference state yields four orthonormal time states depends on the global Hamiltonian spectrum\footnote{It is possible to show that the operator $\hat{V}_C$ spans an orthonormal basis only if $\alpha$ can be written as a ratio of odd and even numbers. We recover the case of evenly-spaced eigenvalues for $\alpha = 1/2 \,,\, 2$ (see Appendix~\ref{app:Res}).}. 
As a consequence, the existence of a discrete resolution of the identity in terms of global time states is not always guaranteed.

In any case, by construction we inherit the resolutions of the identity in terms of local time states and, as shown in Appendix~\ref{app:Res}, it is always possible to find an integral resolution of the identity in terms of global time states. Hence, we recover Eq.s~(\ref{eq:historystatemultiple}-\ref{eq:conditional_Sch}) by inserting these into the Schmidt decomposition of $\ket{\Psi}$. In particular, one gets
\begin{equation}
    \begin{split}
        F(\tau_A,\tau_B|t) &= \bra{\tau_A}_A\bra{\tau_B}_B\ket{t}_C\\&=\prod_{I=A,B}\cos(\omega_I(\tau_I-t))
    \end{split}
\end{equation}
and, assuming $\alpha=p/q\in\mathbb{Q}$,
\begin{equation}
\begin{split}
    \iden_C&=\frac{4}{T}\int_0^T \dd t \, \ketbra{t}_C\\
    &=\alpha\left(\frac{2\omega}{\pi}\right)^2\int_0^{\frac{\pi}{\omega}} \dd\tau \int_0^{\frac{\pi}{\alpha\omega}} \dd\tau' \ketbra{\tau}_A \otimes \ketbra{\tau'}_B
\end{split}
\end{equation}
with $T=\pi q/\omega$. Hence, the state
\begin{equation}
    \ket{\Psi} = \sum_{a,b  = \pm} c_{ab} \ket{a,b}_C \ket{\gamma_{ab}}_S
\end{equation}
of $U$ reads as
\begin{equation}
    \ket{\Psi}=\frac{2}{T}\int_0^T \dd t \, \ket{t}_C\ket{\psi(t)}_S
\end{equation}
from the perspective of the global clock $C$, where we used $a(t)=1/2$ and
\begin{equation}
    \ket{\psi(t)}_{S}=\sum_{a,b} c_{ab} \, e^{i\omega(a+\alpha b)t}\ket{\gamma_{ab}}_S \,.
\end{equation} 
Instead, from the perspective of the local clock $A$ the history state reads as
\begin{equation}
    \ket{\Psi}=\frac{\sqrt{2}\omega}{\pi}\int_0^{\frac{\pi}{\omega}}\dd\tau\,\ket{\tau}_A\ket{\psi(\tau)}_{U | A}\,,
\end{equation}
where we used $a_A(\tau) = 1/\sqrt{2}$ and $\ket{\psi(\tau)}_{U | A}$ is the normalised conditional state of $B$ and $S$, given by
\begin{equation}
    \ket{\psi(\tau)}_{U|A}= \int_0^{\frac{\pi}{\alpha\omega}} \dd \tau' \, \ket{\tau'}_B \ket{\phi(\tau, \tau')}_S\,,
\end{equation}
where
\begin{equation}
    \ket{\phi(\tau, \tau')}_S = 
    \frac{ 4\sqrt{2} \alpha\omega }{T \pi }\!\! \int_0^T \dd t \, F(\tau,\tau'|t)\ket{\psi(t)}_S \,.
\end{equation}
The explicit expression of this state is elaborate but easy to obtain. The generalisation of this construction to a clock model made up of $N$ components is straightforward.

\subsection{The spin-spin interaction}\label{sec:interactingqubits}

In this framework, an example of the gravitational-like interaction of Section~\ref{sec:4} is given by a spin-spin interaction between the local clocks. In particular, the global clock Hamiltonian \eqref{eq:globalclockhamilt} becomes
\begin{equation}\label{eq:eq:globalclockhamiltqubits}
    \hamilt_C = \hbar \omega \big( \hat{\sigma}_A^x + \alpha \hat{\sigma}_B^x - g \hat{\sigma}_A^x \hat{\sigma}_B^x \big) \,,
\end{equation}
where we introduced $g = \alpha \hbar \omega g_{AB}$.
The Time-Dilated Schrödinger equation from $A$'s perspective is recovered from the general case \eqref{eq:SchrInteractGeneral} as
\begin{equation}\label{eq:Schrgravitationalqubits}
     i \hbar \hat{R}\dv{\tau} \ket{\psi(\tau)}_{U|A} = 
     \left(\alpha\hbar \omega \hat{\sigma}_B^x + \hamilt_S \right) 
     \ket{\psi(\tau)}_{U|A} \,,
\end{equation}
where
\begin{equation}
    \hat{R} = \big( \iden - g \hat{\sigma}_B^x \big)
\end{equation}
is the redshift operator defined in \eqref{eq:redshiftopgeneral}. By the  characteristic polynomial
\begin{equation}\label{eq:eigenvaluesR}
    \det(\hat{R} - \epsilon \iden) = 0 \implies \epsilon =
    \begin{cases}
        1 + g \, \text{ on } \ket{-}_B
        \\
        1 - g \, \text{ on } \ket{+}_B
    \end{cases}
    \,,
\end{equation}
we distinguish two cases:  $|g|\neq 1$, for which $\hat{R}$ has rank two and is invertible, and $|g| = 1$, for which $\hat{R}$ is not invertible. 

In the first case, assuming the spectral radius $\rho(g \hat{\sigma}_B^x)$ to be smaller than one, i.e. a bound on the magnitude of the interaction term such that $|g| < 1 $, the geometric series in Eq.~\eqref{eq:geometric} gives
\begin{equation}
    \hat{R}^{-1} = \frac{\iden + g \hat{\sigma}_B^x}{1 - g^2} \,.
\end{equation}
We notice that the same expression for $\hat{R}^{-1}$ holds when $|g| > 1$ and,  inserting this in both sides of the Time-Dilated Schrödinger equation \eqref{eq:Schrgravitationalqubits}, we obtain
\begin{multline}\label{eq:TiDITqubits}
    i \hbar (1 - g^2) \dv{\tau} \ket{\psi(\tau)}_{U|A}
    = \big(\alpha g \hbar \omega + \alpha \hbar \omega \hat{\sigma}_B^x +
    \\
    + \hamilt_S + g \hat{\sigma}_B^x \, \hamilt_S \big) \ket{\psi(\tau)}_{U|A}
     \,.
\end{multline}
This equation provides a simple example of the two main features of this framework: the quantum time-dilation effect and the TiDIT mechanism. 

First, we focus on the quantum time-dilation effect. We can see this by neglecting the free evolution of the system $S$, i.e. setting $\hamilt_S = 0$. The dynamics of $B$ is then described by the relation
\begin{equation}
     i \hbar \dv{\tau} \ket{\psi(\tau)}_{U|A} = \hbar \omega_B \,
     \frac{g + \hat{\sigma}_B^x}{1 - g^2} \,
     \ket{\psi(\tau)}_{U|A} \,.
\end{equation}
Up to a constant energy shift, the non-interacting description of the dynamics of $B$ can be recovered via the coordinate transformation $\dd\tau = \big( 1 - g^2 \big) \dd t $.
Thus, in the presence of this pairwise spin-spin interaction - proportional to the subsystems energy operators - the time interval $\dd \tau$ elapsed from the perspective of $A$ is contracted or expanded, depending on $|g|$, compared to the time interval $\dd t$ in the non-interacting case. 
Also, when $|g|$ is proportional to the reciprocal spatial distance between the two subsystems, the spin-spin interaction is equivalent to a gravitational interaction, and this effect is consistent with the notion of gravitational time-dilation in general relativity. It can thus be read as a quantum time-dilation effect.

On the other hand, relaxing the condition $\hamilt_S = 0$ and neglecting the free dynamics of the local clock $B$, i.e. setting $\omega_B \simeq 0$\footnote{Notice that the condition $ \omega_B = \alpha w \simeq 0 $ implies $|g| \simeq 0$ unless $g_{AB} \propto 1/\omega_B$ and $g_{AB} \omega_B$ is not negligible.}, after the coordinate transformation we obtain
\begin{equation}
     i \hbar \dv{t} \ket{\psi(t)}_{U|A} 
     = \big(\iden + g \hat{\sigma}_B^x \, 
     \big) \hamilt_S
     \ket{\psi(t)}_{U|A} \,.
\end{equation}
From $A$'s perspective, the system $S$ is now coupled with the system $B$. In particular, comparing the new interaction term $+ g \hat{\sigma}_B^x \hamilt_S $ with the one introduced in the global clock Hamiltonian~\eqref{eq:eq:globalclockhamiltqubits} as $ - g \hat{\sigma}_B^x \hamilt_A $, the adopted time-dilated clock description thus transfers its interactions to the other systems, even if originally non-interacting.
This is an example of what we described as the TiDIT mechanism.

Finally, let us focus on the case where the redshift operator is not invertible, i.e. $|g| = 1$, expressing the conditional Schrödinger state for the rest of the Universe in the $\{\,\ket{+}_B \,,\,\ket{-}_B \,\}$ basis as
\begin{multline}
    \ket{\psi(\tau)}_{U|A} = 
    a_+(\tau)\ket{+}_B\ket{\phi_+(\tau)}_S 
    \\ + a_-(\tau) \ket{-}_B\ket{\phi_-(\tau)}_S \,,
\end{multline}
where $a_\pm(\tau)$ are the appropriate normalization coefficients. The eigenvalues of the redshift operator~\eqref{eq:eigenvaluesR} associated with the first and the second term on the right-hand side are $2$ and $0$ respectively. 
As a consequence, substituting this expression into the Time-Dilated Schrödinger equation~\eqref{eq:Schrgravitationalqubits} we get
\begin{equation}
     \hat{R} \, \dv{\tau} \ket{\psi(\tau)}_{U|A} = 
     2 \, \dv{\tau} a_-(\tau) \ket{-}_B\ket{\phi_-(\tau)}_S \,;
\end{equation}
while the dynamics of the $\ket{-}_B$ subspace are still described by a Schrödinger equation, the other component obeys the constraint
\begin{equation}\label{eq:degeneracyqubits}
    \left( \alpha\hbar \omega + \hamilt_S \right) \ket{\phi_+(\tau)}_S = 0 \,.
\end{equation}
The state $\ket{\phi_+(\tau)}_S = \ket{-\alpha\hbar w}_S$ is thus a stationary state, and the history state corresponding to this case can be written as
\begin{multline}\label{eq:frozensolution}
    \ket{\Psi} = \int \dd \mu(\tau) \, a_-(\tau) \ket{\tau}_A \ket{-}_B\ket{\phi_-(\tau)}_S + 
    \\
    + \left( \int \dd \mu(\tau) \, a_+(\tau)  \ket{\tau}_A \right) \ket{+}_B\ket{-\alpha\hbar w}_S \,, 
\end{multline}
i.e. as a superposition of an entangled (dynamical) history for the subspace associated with the $\ket{-}_B$ state, and a disentangled (stationary) history for the one associated with the $\ket{+}_B$ state. 

\begin{figure}
    \centering   \includegraphics[width=0.4\textwidth]{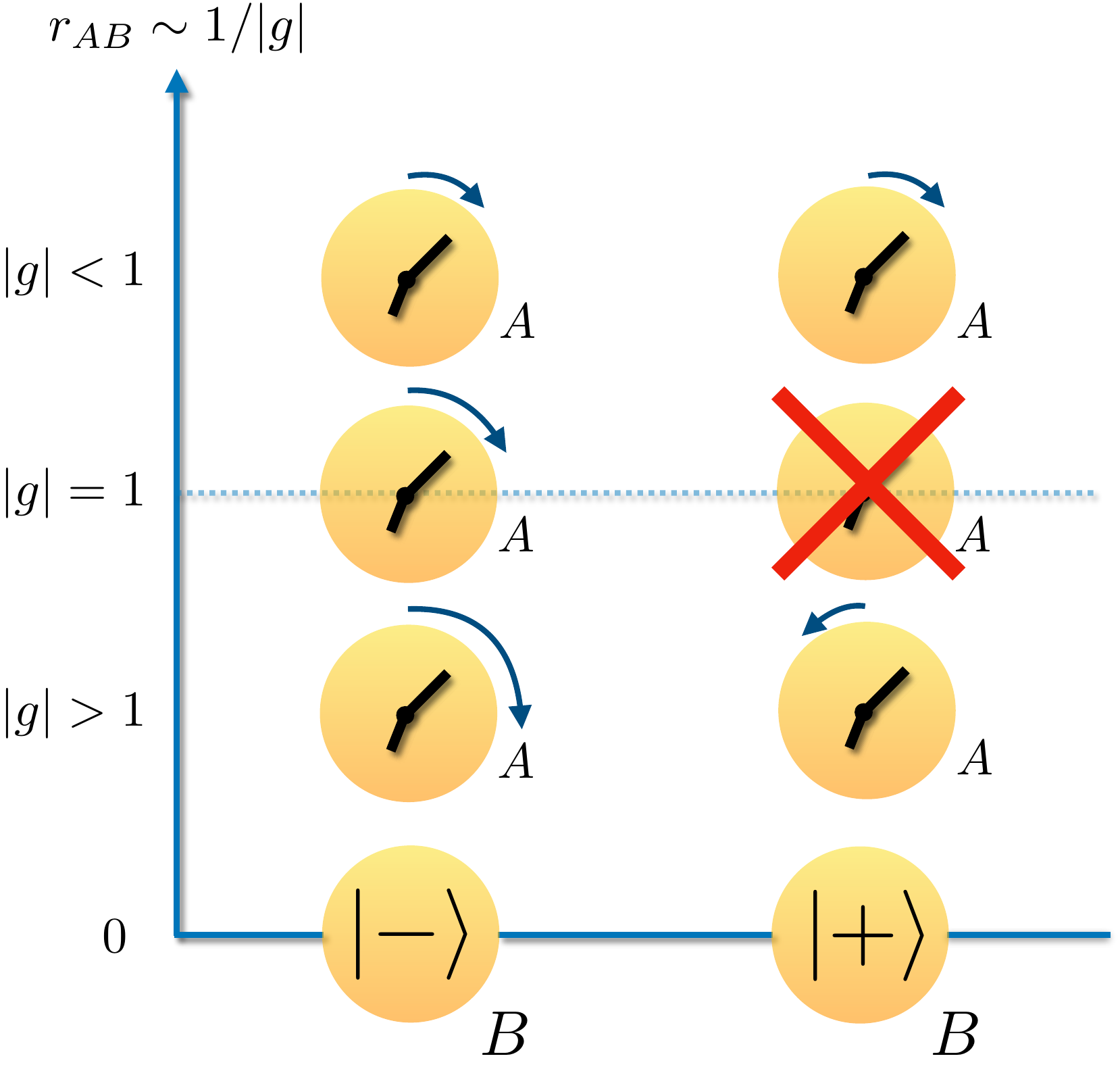}
    \caption{Time evolution as a function of the interaction strength $|g|$ and $\hat{R}$'s eigenstates. Representing the reciprocal of the coupling strength $1/|g|$ as the distance $r_{AB}$ between the local clocks $A$ and $B$, the time evolution described by the clock $A$ and associated with the $B$'s minus state is time-dilated as a function of $|g|$ but always proceeds clockwise. On the other hand, the time evolution associated with the $B$'s plus state exhibits three different behaviours. In particular, evolution is stuck at the non-invertibility point $|g|=1$ and reversed for $|g|>1$.}
    \label{fig:horizon}
\end{figure}

In conclusion, the non-invertibility of the redshift operator in $|g| = 1$ represents a transition between two distinct dynamical behaviours: when $|g| < 1$, both the eigenvalues in the spectrum~\eqref{eq:eigenvaluesR} are positive and the two subspaces evolve with the same (positive time) parameter; instead, when $|g| > 1$, one of the eigenvalues changes sign, describing a reversed dynamical behaviour (negative time) for one of the subspaces.
The transition takes place in $|g| = 1$, where the global clock Hamiltonian becomes degenerate. This effect is represented in Fig.~\ref{fig:horizon}. 

An example similar to the one presented above has also been discussed in Ref.s~\cite{SmithM19,Rijavec23}, where the dynamics induced by the spin-spin interaction are analyzed. Compared with these works, the novelties of our discussion are the following. In contrast with Ref.~\cite{SmithM19}, where the authors adopt a more general interaction (including other spin-spin components) and offer a thorough perturbative analysis of the dynamics, our approach provides a non-perturbative description of the gravitationally interacting qubit clock, resulting in a dynamical behaviour akin to that described in Ref.~\cite{Rijavec23} by an ideal clock. Furthermore, introducing a redshift operator on the same line of Ref.~\cite{CastroRuizEtAl17}, our construction leads to the interpretation of this behaviour in terms of a quantum mechanical time dilation effect and the observation of the herein presented TiDIT mechanism.

Despite the mathematically involved expression for the inverse of the redshift operator, the generalization of this construction to a $N$-spins clock model with spin-spin interactions proportional to the free-spin Hamiltonians is straightforward, while the two qubits model already captures the relevant physics introduced by this finite-dimensional framework.

\section{Discussion and outlook}\label{sec:Discussion}

In this work, we used the Page and Wootters mechanism \cite{PageW83,Wootters84} to describe a theory of quantum time for finite-dimensional quantum systems. In analogy with the well-known ideal (read, infinite-dimensional) case, our construction describes time as the intrinsic quantum correlations of a constrained bipartite system. In particular, the constraint is implemented by a Hamiltonian operator whose action on the subspace of interest is null; those states satisfying the constraint equation and presenting entanglement between the two parts of the system, called \textit{clock} and \textit{the rest}, enable an emergent description of time. This is achieved by using a common label for the clock states and those of the remaining subsystem. In the ideal case, this construction is widely known and studied in the literature and has inspired several recent elaborations aimed at solving related interpretative and practical problems~\cite{GiovannettiEtAl15, HoehnEtAl21, HoehnEtAl20, HausmannEtAl23}. On the contrary, the finite-dimensional version of the Page and Wootters construction has received little attention and is less explored in existing research. 
The use of a POVM for the description of quantum time with a finite-dimensional clock was first discussed in Ref.~\cite{Pegg98}, but the first implementation in this context with a qubit clock~\cite{SmithM19}, and its further developments~\cite{FavalliS22,FavalliS23} are still relatively new. Building on that, our construction fills this gap and offers several additional advantages. First, its simplicity permitted the analysis of multipartite clocks, making it possible to describe a new phenomenon called TiDIT (namely, the appearance of an effective interaction term between a system and its multipartite clock when the flow of time is parametrized by one of the subclocks only), and the emergence of gravitational-like effects in terms of a so-called redshift operator. Second, our model provides a controllable framework for speculative theoretical work in fundamental many-body quantum physics and low-energy quantum gravity, as well as a formal setup suitable for experimental implementations. These advantages of our model and potential future developments are discussed in more detail below.

First, let us briefly expand on the new results we found thanks to the simplicity of our finite-dimensional approach. Considering a system composed of $N$ (possibly interacting) clocks and a system to which these clocks provide time, the Page and Wootters mechanism can be directly applied when the ``local'' clocks act together as a unique ``global'' clock for the system. Moreover, one can consider the case where one of the subclocks provides time for all the rest: in this case, a non-perturbative effective Hamiltonian for the latter emerges. An explicit expression for this new Hamiltonian can be found when choosing the interaction between the local clocks to commute with their free evolutions, a property we achieved using the so-called gravitational-like interactions. This unravelling reveals the curious appearance of interactions between all parts of the evolving system, including a coupling between the initially non-interacting system and all clocks. This result is derived from the Time-Dilated Schrödinger equation via a newly identified mechanism, dubbed the Time-Dilation induced Interaction Transfer (TiDIT) mechanism, that introduces these interactions. Because the TiDIT mechanism requires the invertibility of a so-called redshift operator $\hat{R}$, finding this result is not always possible. Interestingly, a study of cases where the mechanism is unavailable shows the peculiar freezing of some subspaces of the evolving Hilbert space in correspondence with specific values of the subclock's couplings. As we will see in the next paragraph, the resulting description is consistent with the behaviour of quantum systems in the presence of gravitational fields and directly leads to exciting speculations about time near event horizons. Overall, the potential of our formulation becomes evident when considering a two-qubit clock model, where the formal structure of the equations is particularly manageable and provides the starting point for studying the infinite-dimensional limit and theoretical and technological implementations for exploring gravity at the intersection with quantum physics.

While a discussion of whether this framework is really able to capture some fundamental aspects of the quantum nature of gravity is beyond the scope of this paper, certain features can be discussed based on the correspondence between the redshift operator and the square root of the $g_{00}$ component of the metric tensor in a Schwarzschild space-time, discussed in Ref.~\cite{CastroRuizEtAl20}. 
Indeed, assuming the qubit clocks in Section~\ref{sec:5} are placed in a $3$-dimensional space at fixed reciprocal distance $r$, the redshift operator $\hat{R}$ can be written in the form $\iden - \hat{r}_S / r $, where $\hat{r}_S:= G \hamilt / c^4 = r g \hat{\sigma}^x $, hence encoding the Schwarzschild radius of the source qubit with gravitational constant $G$. In consequence, this model can potentially be suitable for describing the dynamics of gravitationally interacting finite-dimensional quantum systems in the weak-field approximation; this is also consistent with the description arising in the case of a macroscopic source of the gravitational field~\cite{FavalliS23}, and may be used to implement a description of a superposition of gravitational fields~\cite{Giacomini21}. Moreover, this fact opens up fascinating speculations on the quantum description of an event horizon, as the non-invertibility point of $\hat{R}$ is given by the condition $r = r_S^*$, with $r_S^*$ in the spectrum of $\hat{r}_S$, and a description of the dynamics close to this point can be recovered via the TiDIT mechanism. 

Concerning potential technological implementations, our description has the additional advantage of allowing the design of simple finite-dimensional analogue experimental platforms on which to test selected quantum-gravity ideas. Indeed, the implementation of our framework with qubit systems and current quantum technology may pave the way for new avenues of inquiry into the theory of computation with indefinite causal structures~\cite{Hardy09} and quantum algorithms in a space-time superposition~\cite{Shmueli24}.
This still requires the development of the appropriate laboratory measurement scheme for the description of internal dynamical evolution, which represents a further step in the establishment of our framework and will be addressed in future works. 

Finally, we notice that our construction leads to distinguishing features with respect to the ideal case, i.e. beyond our finite-dimensional Time-Dilated Schrödinger equation.   In the derivation, a normalization coefficient $a(t)$ naturally arises via the Schmidt decomposition of the considered stationary state of the Universe, by requiring that the conditional state of the system of interest is always normalized. Furthermore, this coefficient can only be constant in $t$ if the constraint is of the type discussed in Sec.~\ref{sec:4}. A non-constant coefficient and the corresponding contribution to the dynamical laws are expected if a different constraint is assumed, e.g. a constraint containing the so-called \textit{Time operator}~\cite{GiovannettiEtAl15}. In addition, the generalization of the TiDIT mechanism to the ideal case may involve divergences that are difficult to handle, even invoking perturbation theory.
Thus, the infinite-dimensional limit still requires further clarification and will be discussed in future works. 

\section*{Acknowledgments}
The authors are deeply grateful to Flaminia Giacomini for the critical reading of the manuscript and to Sabrina Maniscalco for valuable input during the initial stages of this project. D.C. also acknowledges valuable confrontation with Philipp Höhn and thanks C. De Rosa for insightful remarks.  N.P. acknowledges financial support from the Magnus Ehrnrooth Foundation and the Academy of Finland via the Centre of Excellence program (Project No. 336810 and Project No. 336814). J.Y.M. and M.L.C. were supported by the European Social Fund REACT EU through the Italian national program PON 2014-2020, DM MUR 1062/2021. M.L.C. acknowledges support from the National Centre on HPC, Big Data and Quantum Computing—SPOKE 10 (Quantum Computing) and received funding from the European Union Next-GenerationEU—National Recovery and Resilience Plan (NRRP)—MISSION 4 COMPONENT 2, INVESTMENT N. 1.4—CUP N. I53C22000690001. This research has received funding from the European Union’s Digital Europe Programme DIGIQ under grant agreement no. 101084035. M.L.C. also acknowledges support from the project PRA2022202398 “IMAGINATION”.

%%%%%%%%%%%%%%%%%%%%%%%%%%%%%%%%%%%%%%%%%%%
% DOCUMENT ENDING
%%%%%%%%%%%%%%%%%%%%%%%%%%%%%%%%%%%%%%%%%%%

\appendix

\section{Resolution of the Identity}
\label{app:Res}

Here we show how to derive a resolution of the Identity for the clock system $C$ in terms of time states. Notice that time states are generally non-orthogonal when dealing with finite-dimensional quantum systems. In other words, the transition amplitude between the time states $\ket{t}$ and $\ket{s}$, given by
\begin{equation}\label{eq:overlap_function}
    \braket{s}{t} = \frac{1}{d_C}\sum_{k=0}^{d_C - 1} e^{i w_k (s - t)} \,,
\end{equation}
is not a delta over $t$ and depends on the spectrum of the clock Hamiltonian. 

Let us consider a Hamiltonian with evenly spaced eigenvalues in the form
\begin{equation}\label{eq:clockspectrum}
    w_k := w_0 + k \, \frac{2\pi}{T} \,, \text{ with } k = 0, ..., d_c - 1 \,.
\end{equation}
Introducing the discrete-time parameter $t_n = n \, T / d_C $ with $n \in \mathds{Z}$, the transition amplitude between the time states $\ket{t_a}$ and $\ket{t_b}$ becomes
\begin{equation}
    \bra{t_b}\ket{t_a} 
    = \frac{1}{d_C} e^{i w_0 \frac{T(b - a)}{d_C}} \sum_{k=0}^{d_C - 1} e^{2\pi i \frac{k(b - a)}{d_C}} 
    = \delta_{ab} \,,
\end{equation}
where $a,b=0,...,d_C - 1$. 
We can extract an orthonormal set of time states $\{ \ket{t_n} \}_{n=0, ..., d_C - 1} $ providing a basis for $\hilbert_C$ and a resolution of the Identity as
\begin{equation}\label{eq:resolution_discrete}
    \iden_C = \sum_{n=0}^{d_C - 1} \ketbra{t_n} 
    = \frac{1}{d_C} \sum_{n=0}^{d_C - 1} \ketbra{\bar{t}_n}
    \,,
\end{equation}
where $\ket{\bar{t}_n} = \sqrt{d_C}\ket{t_n}$ is a non-normalized time state satisfying the relation $\bra{w_k}\ket{\bar{t}_n} = e^{-i w_k t_n}$.

More generally, let us consider a Hamiltonian with eigenvalues having rational ratios - up to a constant shift $w_0$, in the form
\begin{equation}
    w_k = w_0 + r_k \, \frac{2 \pi}{T} \,, \text{ with } k = 0, ..., d_c - 1 \,,
\end{equation}
in which
\begin{equation}\label{eq:ratios}
    \frac{r_k}{r_1} := \frac{A_k}{B_k} = \frac{w_k - w_0}{w_1 - w_0} 
    \quad \text{ and } 
    \quad T = \frac{2\pi r_1}{w_1 - w_0} \,, 
\end{equation}
where $A_k, B_k \in \mathds{N}$ have no common factors and $r_1$ is the lowest common multiple of the values of $B_k$~\cite{Pegg98,FavalliS23}.
The special case of evenly-spaced eigenvalues is recovered when $r_k = k$.
Let us redefine the discrete-time parameter as $t_n = n \, T/N $ with $N > \max\{r_k\}$.
As a consequence, the time states associated with different values of $t_n$ are generally non-orthogonal, but the ones ranging from $n = 0$ to $N - 1$ provide an over-complete basis and a resolution of the Identity~\cite{Pegg98} as
\begin{equation}\label{eq:resolution_discrete_overcomp}
    \iden_C = \frac{d_C}{N} \sum_{n = 0}^{N-1} \ketbra{t_n} = \frac{1}{N} \sum_{n = 0}^{N-1} \ketbra{\bar{t}_n} \,.
\end{equation}
Multiplying and dividing by the period $T$ and sending $\Delta t := T/N$ to zero, the sums become integrals and the expression above becomes
\begin{equation}\label{eq:resolution_continu}
    \iden_C = \frac{d_C}{T} \int_{0}^{T} \dd t \; \ketbra{t} 
    = \frac{1}{T} \int_{0}^{T} \dd t \; \ketbra{\,\bar{t}\,} \,.
\end{equation}
When $r_k = k$ and $N = d_C$, the resolution of the Identity in Eq.~\eqref{eq:resolution_discrete_overcomp} reduces to Eq.~\eqref{eq:resolution_discrete}, and the limit $\Delta t \to 0$ corresponds to the infinite-dimensional limit, i.e. $d_C \to + \infty$. In this case, the non-normalized time states in the integral expression~\eqref{eq:resolution_continu} become orthogonal (i.e. setting $k \in \mathds{Z}$ in the clock spectrum~\eqref{eq:clockspectrum}, the transition amplitude~\eqref{eq:overlap_function} between non-normalized time states becomes a delta function) and the spectrum~\eqref{eq:clockspectrum} becomes equivalent to the one of a free particle in a box of length $T$. 

Finally, this result can be generalized to the case of irrational eigenvalue ratios pointing out that any real number can be approximated with arbitrary precision by a rational number. Hence, Eq.~\eqref{eq:ratios} can still be used to approximate irrational ratios up to a certain precision, leading to a very large denominator $r_1$ and a very large period $T$. The limit of $T \to + \infty$ then corresponds to the case of non-commensurable eigenfrequencies~\cite{Pegg98} or continuous energy spectrum.
More generally, we thus introduce the expression
\begin{equation}\label{eq:resolution_app}
    \iden_C = \int \dd \mu(t) \ketbra{t} \,,
\end{equation}
where the integration measure $\dd \mu(t)$ compactly represents all of the previous cases. 

\section{The normalization coefficient}
\label{app:lambda}

The conditional Schrödinger state $\ket{\psi(t)}_S$ is well-defined for all $t$ s.t. $a(t)\neq 0$. The physical interpretation of this coefficient is described in Ref.~\cite{Pegg98} employing the resolution of the Identity~\eqref{eq:resolution_continu}. This expression satisfies the statistical properties of a POVM generated by the infinitesimal operator $\frac{d_C}{T} \ketbra{t}_C \dd t $ as
\begin{multline}
    \braket{\Psi} = \bra{\Psi}\iden_C\ket{\Psi} = \frac{d_C}{T} \int_{0}^{T} \dd t \; \norm{\bra{t}_C\ket{\Psi}}^2 =
    \\
    = \int_{0}^{T} \dd t \; \frac{d_C}{T} a^2(t) := \int_{0}^{T} \dd t \; \Pr(t) = 1\,.
\end{multline}
Thus, the probability of finding the composite system in a state associated with the time parameter $t$ is proportional to $a^2(t)$. Conversely, when $a(t)$ is constant, the distribution $\Pr(t)$ is homogeneous and the composite state $\ket{\Psi}$ is maximally delocalized in time. 

This is the case when the Schmidt decomposition of $\ket{\Psi}$ is given in terms of clock energy eigenstates. Indeed, an explicit computation gives
\begin{multline}\label{eq:coefficient}
    a^2(t) = \sum_{n=0}^{r-1} \lambda_n |\bra{\phi_n}\ket{t}|^2
    = \sum_{a,b=0}^{d_C -1} c_{ab} \, e^{-i(w_a - w_b)t} 
    \,,
\end{multline}
where 
\begin{equation}
    c_{ab} = \frac{1}{d_C} \sum_{n = 0}^{r -1} \lambda_n  \bra{\phi_n}\ket{w_a}_C\bra{w_{b}}\ket{\phi_n}_C \,.
\end{equation}
When the orthonormal states $\{ \ket{\phi_n} \}$ coincide with the energy eigenstates $\{\ket{w_n}\}$, we have
\begin{equation}
    a^2(t) = \frac{1}{d_C} \sum_{a,b,n} \lambda_n \, \delta_{na}\,\delta_{bn} \, e^{-i(w_a - w_b)t} = \frac{1}{d_C} \,,
\end{equation}
where we employed the normalization condition
\begin{equation}
    \braket{\Psi} = \sum_{n=0}^{r-1} \lambda_n = 1 \,.
\end{equation}
Thus, the probability distribution of the time states is homogeneous and is given by $1/T$.
The same is true when $\ket{\Psi}$ is an eigenstate of the non-interacting Hamiltonian~\eqref{eq:Hamiltonian_CG} or in the presence of a gravitational-like interaction~\eqref{eq:gravitational-likeinteraction}. Indeed, it is easy to see that the non-normalized state $\bra{t}_C\ket{\Psi}$ evolves unitarily in $t$, and thus $a(t)$ must be a constant.

\bibliographystyle{unsrt} 

\end{document}